\documentclass[11pt,a4paper]{article}

\usepackage{graphicx}
\usepackage{amssymb}

\usepackage{amsfonts}
\usepackage{geometry}


\begin{document}

\title{All Mutually Unbiased Bases\\ in\\ Dimensions Two to Five}
\author{{\normalsize Stephen Brierley and Stefan Weigert} \\
{\normalsize Department of Mathematics, University of York}\\
{\normalsize Heslington, UK-York YO10 5DD}\\
\\
{\normalsize Ingemar Bengtsson} \\
{\normalsize Fysikum, Stockholms Universitet}\\
{\normalsize S-106 91 Stockholm, Sweden}\\
\\
\texttt{\normalsize sb572@york.ac.uk, slow500@york.ac.uk, ingemar@physto.se}}

\maketitle

\begin{abstract}
All complex Hadamard matrices in dimensions two to five are known. We use
this fact to derive all inequivalent sets of mutually unbiased (MU) bases in
low dimensions. We find a three-parameter family of triples of MU bases in
dimension four and two inequivalent classes of MU triples in dimension five.
We confirm that the complete sets of $(d+1)$ MU bases are unique (up to
equivalence) in dimensions below six, using only elementary arguments for $d$
less than five.
\end{abstract}

\section{Introduction}

Position and momentum of a \textit{classical} non-relativistic particle are
intimately linked since the momentum variable generates spatial
translations. Mathematically, this important relation is embodied in the
structure of the Galilei group. It turns out to be even more fundamental for
a \textit{quantum mechanical} particle, where it takes the form of the
commutation relation of its position and momentum operators.

The associated Heisenberg-Weyl group of phase-space translations continues
to be relevant for quantum systems with only a \textit{finite} number of
orthogonal states, providing a basis of the space $\mathbb{C}^d$. For each
dimension $d \geq 2$, there is a set of unitary operators which give rise to
a discrete equivalent of the Heisenberg-Weyl group \cite%
{schwinger60}. Physicists would expect the state spaces $\mathbb{C}^d$ to be
structurally identical, at least with respect to properties closely related
to the Heisenberg-Weyl group. It thus comes as a surprise that the
Heisenberg-Weyl group allows one to construct $(d+1)$ so-called mutually
unbiased (MU) bases of the space $\mathbb{C}^d$ if $d$ is the power of a
prime number \cite{ivanovic81,wootters+89} while the construction fails in
all other 'composite' dimensions. No other successful method to construct $%
(d+1)$ MU bases in all dimensions is known \cite{archer05, planat+06, roy+07}%
.

Given $(d+1)$ orthonormal bases in the space $\mathbb{C}^{d}$, they are 
\textit{mutually unbiased} if the moduli of the scalar products among the $%
d(d+1)$ basis vectors take these values: 
\begin{equation}  \label{MUB conditions}
\left\vert {\langle \psi _{j}^{b}|}{\psi _{j^{\prime }}^{b^{\prime }}
\rangle }\right\vert =\left\{ 
\begin{array}{ll}
\delta_{jj^{\prime }} & \quad \mbox{if $b = b^\prime$}\, , \\ 
\frac{1}{\sqrt{d}} & \quad \mbox{if $b \neq b^\prime $}\,,%
\end{array}
\right. 
\end{equation}
where $b,b^{\prime }=0,1,\ldots ,d$. Such \textit{complete} sets of MU bases
are ideally suited to reconstruct quantum states \cite{wootters+89} while
sets of up to $(d+1)$ MU bases have applications in quantum cryptography 
\cite{cerf+02, brierley09} and in the solution of the Mean King's problem \cite%
{aharonov+01}, for example.

The methods to construct complete sets of MU bases typically deal with all
prime or prime-power dimensions simultaneously. They either make use of the
Heisenberg-Weyl group \cite{bandyopadhyay+2001}, exploit identities from
number theory \cite{wootters+89,klappenecker+04}, or they are couched in the language of
finite fields \cite{wootters+89,durt04}. All these methods are \textit{%
constructive} and effectively lead to the same bases. The existence of
other, inequivalent sets of complete MU bases remains unclear.

In this paper, we choose a different method to study MU bases in
dimensions two to five. It allows us to directly conclude that the
corresponding complete sets of $(d+1)$ MU bases are unique (up to some
irrelevant equivalence, cf. below). What is more, we are also able to
exhaustively list all inequivalent classes of $d$ or less MU bases. This
approach is attractive because it uses elementary methods only, except in
dimension five. In order to present the details of this method we need to
briefly discuss the relation between MU bases and complex Hadamard matrices.

\subsection*{MU bases and complex Hadamard matrices}

Each MU basis in the space $\mathbb{C}^d$ consists of $d$ orthogonal unit
vectors which, collectively, will be thought of as a unitary $d \times d$
matrix. Two (or more) MU bases thus correspond to two (or more) unitary
matrices, one of which can always be mapped to the identity $I$ of the space 
$\mathbb{C}^{d}$, using an overall unitary transformation. It then follows from the
conditions (\ref{MUB conditions}) that the remaining unitary matrices must
be complex Hadamard matrices: the moduli of all their matrix elements equal $%
1/\sqrt{d}$. This representation of MU bases links their classification to
the classification of complex Hadamard matrices \cite{tadej+06,Hadamardsonline}.

It is, for example, possible to list all pairs of MU bases $\{I,H\}$ in $\mathbb{C}^d$ once all complex Hadamard matrices are known. Using
this observation as a starting point, we will extend the classification from
pairs to sets of $r\leq (d+1)$ MU bases. As complex Hadamard matrices have been
classified for $d\leq 5$, we expect to obtain an exhaustive list of sets of $%
r$ MU bases in these low dimensions.

The approach we take is inspired by recent works aimed at dimension six \cite%
{grassl04,brierley+09,Jaming+09}. Starting with a pair of MU\ bases $\{I,H\}$
we will find \emph{all} vectors $v$ which are MU to $I$ and $H$. Only these
vectors represent candidates to form additional bases and by analysing their
inner products we will be able to obtain \emph{all} MU bases in low
dimensions.

In dimension six it is not trivial to determine all candidate vectors. It
becomes necessary to use Gr\"obner bases \cite{brierley+09} or to discretise
the underlying space \cite{Jaming+09}. For $d \leq 5$, however, we find
that elementary properties of the complex plane are sufficient to solve (most of) the relevant
equations in closed form.

The task to find all MU bases is complicated by the fact that,
actually, many sets of apparently different MU bases are identical to each
other. For the desired classification, it is sufficient to enumerate all 
\emph{dephased} sets of $(r+1)$ MU bases. This \emph{standard form} \cite%
{tadej+06} is given by 
\begin{equation}  \label{standardform}
\{ I, H_{1},\ldots, H_{r}\}\, , \quad r \in \{1, \dots ,d\} \, ,
\end{equation}
where $I$ is the identity in $\mathbb{C}^d$ and the other matrices are
complex Hadamard matrices of a particular form: the components of the first
column of the matrix $H_{1}$ are given by $1/\sqrt{d}$, and the first row of
each Hadamard matrix $H_{\rho}, 1 \leq \rho \leq r$ has entries $1/\sqrt{d}$ only (see
Eqs. (\ref{f5}) and (\ref{d=5 complete set}) for an explicit example with $%
d=5$). The possibility of dephasing is based on the notion of \emph{%
equivalence classes} for MU bases, explained in more detail in Appendix A.
Roughly speaking, two MU bases are equivalent to each other if one can be
obtained from the other by changing the overall phases of individual vectors
and by permuting them.

The results of this paper have been arranged as follows. In Sec. \ref{dim 23}
we deal with dimensions two and three. The complete list of sets of MU bases in
dimension four is derived in Sec. \ref{dim 4}. Then, all sets of MU bases of $\mathbb{C}%
^5$ are constructed, and in Sec. \ref{concl} we summarize and discuss our
results.

\section{Dimensions $d=2$ and $d=3$ \label{dim 23}}

In this section, we construct all sets of MU bases in dimensions two and three using
only simple properties of the complex plane. The direct approach to construct all MU bases for $d=4$
in Sec. \ref{dim 4} will be based on similar arguments.

\subsection{Dimension $d=2$}

The matrices consisting of the eigenvectors of the Heisenberg-Weyl operators
form a set of three MU bases in dimension two which are unique up to the
equivalences specified in Appendix A. 
We present a simple
proof of this well-known fact.

Let us begin by noting that there is only one dephased complex Hadamard
matrix in $d=2$ (up to equivalences), the discrete $(2\times2)$ Fourier matrix 
\begin{equation}
F_{2}=\frac{1}{\sqrt{2}}\left(%
\begin{array}{cc}
1 & 1 \\ 
1 & -1%
\end{array}%
\right)\,.  \label{2d fourier}
\end{equation}
A vector $v\in\mathbb{C}^{2}$ is MU to the standard basis $I$ (constructed
from the eigenstates of the $z$-component of a spin $1/2$) if its components
have modulus $1/\sqrt{d}$. Applying the transformation given in Eq. (\ref%
{diagunitaries}), the dephased form of such a a vector reads $%
v=(1,e^{i\alpha})^{T}/\sqrt{2}$, with a real parameter $\alpha\in\left[0,2\pi%
\right]$. The vector $v$ is MU to the columns of $F_{2}$ if the phase $\alpha
$ satisfies two conditions, 
\begin{equation}
\left\vert 1\pm e^{i\alpha}\right\vert =\sqrt{2}\,.  \label{d=2 constraints}
\end{equation}
These equations hold simultaneously only if $e^{i\alpha}=\pm i$. Thus, there
are only two vectors which are MU to both $I$ and $F_{2}$, given by $%
v_{\pm}=\left(1,\pm i\right)^{T}/\sqrt{2}$. Since this is a pair of
orthogonal vectors, they form a Hadamard matrix $H_{2}=(v_{+}|v_{-})$ and,
therefore, the three sets 
\begin{equation}
\{I\}\,,\{I,F_{2}\}\,,\{I,F_{2},H_{2}\}  \label{d=2 triples}
\end{equation}
represent all (equivalence classes of) one, two, or three MU bases in
dimension two.

\subsection{Dimension $d=3$ \label{dim 3}}

In dimension three there is also only one dephased complex Hadamard matrix up to equivalence. It is given by the $(3\times3)$ discrete Fourier matrix%
\begin{equation}
F_{3}=\frac{1}{\sqrt{3}}\left(%
\begin{array}{ccc}
1 & 1 & 1 \\ 
1 & \omega & \omega^{2} \\ 
1 & \omega^{2} & \omega%
\end{array}%
\right)\,,  \label{F3}
\end{equation}
defining $\omega=e^{2\pi i/3}.$ Again, we search for dephased vectors $%
v=(1,e^{i\alpha},e^{i\beta})^{T}/\sqrt{3}$, $0\leq\alpha,\beta\leq2\pi$ ,
which are MU with respect to the matrix $F_{3}$. This leads to the following
three conditions 
\begin{eqnarray}
\left\vert 1+e^{i\alpha}+e^{i\beta}\right\vert & = & \sqrt{3}\,,  \nonumber
\\
\left\vert 1+\omega e^{i\alpha}+\omega^{2}e^{i\beta}\right\vert & = & \sqrt{3%
}\,,  \label{d=3 constraints} \\
\left\vert 1+\omega^{2}e^{i\alpha}+\omega e^{i\beta}\right\vert & = & \sqrt{3%
}\,.  \nonumber
\end{eqnarray}

\begin{figure}[ht]
\begin{center}
\includegraphics[width=0.5\textwidth]{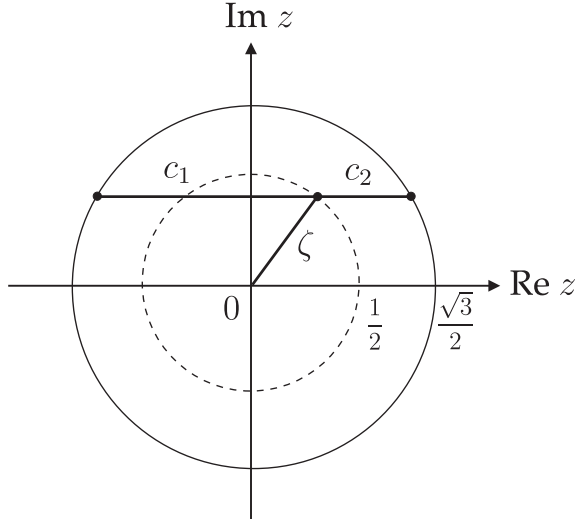} 
\end{center}
\caption{Plot of Eqs. (\ref{d=3 geometric constraint}) in the
complex $z$-plane. The real numbers $c_1$ and $c_2$ each represent one
of the three numbers $\cos(\protect\alpha/2)$ and $\cos(\protect\alpha/2\pm2%
\protect\pi/3)$; it follows that at least two of these three expressions
must be equal.}
\label{threeplane}
\end{figure}

Removing an overall factor of $e^{i\alpha/2},$ they can be rewritten 
\begin{eqnarray}
\left\vert \zeta +\cos\frac{\alpha}{2}\right\vert & = & \frac{\sqrt{3}}{2}\,,
\nonumber \\
\left\vert \zeta +\cos\left(\frac{\alpha}{2}\pm\frac{2\pi}{3}%
\right)\right\vert & = & \frac{\sqrt{3}}{2}\,,
\label{d=3 geometric constraint}
\end{eqnarray}
where $2\zeta =e^{i(\beta-\alpha/2)}$. By considering a plot in the complex
plane, Fig. \ref{threeplane}, we see that these three equations hold
simultaneously only if two of the cosine terms are equal. This implies that
the only possible values of the parameter $\alpha$ are $0,\pi/3,$ or $2\pi/3$%
, leading to the requirement $\pm1/2=\cos\beta$. Consequently, the Eqs. (\ref%
{d=3 constraints}) have exactly six solutions which give rise to vectors 
\begin{eqnarray}  \label{six MU vectors}
v_{1} & \propto & \left(%
\begin{array}{c}
1 \\ 
\omega \\ 
\omega%
\end{array}%
\right)\,,\quad v_{2} \propto \left(%
\begin{array}{c}
1 \\ 
\omega^{2} \\ 
1%
\end{array}%
\right)\,,\quad v_{3} \propto \left(%
\begin{array}{c}
1 \\ 
1 \\ 
\omega^{2}%
\end{array}%
\right)\,,  \nonumber \\
v_{4} & \propto & \left(%
\begin{array}{c}
1 \\ 
\omega^{2} \\ 
\omega^{2}%
\end{array}%
\right)\,,\quad v_{5} \propto \left(%
\begin{array}{c}
1 \\ 
\omega \\ 
1%
\end{array}%
\right)\,,\quad v_{6} \propto \left(%
\begin{array}{c}
1 \\ 
1 \\ 
\omega%
\end{array}%
\right) \,.
\end{eqnarray}

Examining their inner products shows that there is only one way to arrange
them (after normalization) into two orthonormal bases, namely $%
H_{3}^{(1)}=(v_{1}|v_{2}|v_{3})$ and $H_{3}^{(2)}=(v_{4}|v_{5}|v_{6})$. It
is useful to note that one can write 
\begin{equation}  \label{powers of D}
H_{3}^{(1)} = D F_{3} \quad \mbox{ and } \quad H_{3}^{(2)} = D^2 F_{3} \, ,
\end{equation}
where $D={\mbox{diag}(1,\omega,\omega)}$ is a diagonal unitary matrix with
entries identical to the components of the vector $v_1$, i.e. the first
column of $H_3^{(1)}$. The triples obtained from adding either $H_{3}^{(1)}$
or $H_{3}^{(2)}$ to the pair $\left\{ I,F_{3}\right\} $ are equivalent, 
\begin{eqnarray}  \label{one triple only}
\{I,F_{3},H_{3}^{(1)}\} \sim \{D^2ID,D^2F_{3},D^2 H_{3}^{(1)}\} =
\{I,H_{3}^{(2)},F_{3}\} \sim \{I,F_{3},H_{3}^{(2)}\} \, ,
\end{eqnarray}
as follows from first applying the unitary $D^2$ globally from the left,
rephasing the first basis with $D^{-2}\equiv D$, and finally rearranging the
last two bases. We therefore conclude that the sets constitute a complete classification of all sets of MU bases in dimension $d=3$.

\section{Dimension $d=4$ \label{dim 4}}

In dimension $d=4$, a one-parameter family of complex
Hadamard matrices exists, 
\begin{equation}
F_{4}(x)=\frac{1}{2}\left(%
\begin{array}{cccc}
1 & 1 & 1 & 1 \\ 
1 & 1 & -1 & -1 \\ 
1 & -1 & ie^{ix} & -ie^{ix} \\ 
1 & -1 & -ie^{ix} & ie^{ix}%
\end{array}%
\right)\,,\quad x\in[0,\pi]\,.  \label{d=4 Fourier Hadamard}
\end{equation}
It is elementary to show that all $4 \times 4$ complex Hadamard matrices are equivalent to a member of the family $F_{4}(x)$. When $x=0$, the resulting matrix is the discrete Fourier
transform $F_{4}$ on the space $\mathbb{C}^{4}$, with matrix elements given
by $\omega^{jk},j,k=0\ldots3, \omega\equiv i$. The matrix $F_{4}(\pi/2)$ is
equivalent to a direct product of the matrix $F_{2}$ with itself while for
other values of $x$ it can be written as a Hadamard product of $F_{4}$ with
an $x$-dependent matrix.

\subsection{Constructing vectors MU to $F_{4}(x)$ \label{dim 4 construction}}

After dephasing, any vector MU to the standard basis takes the form $%
v=(1,e^{i\alpha^{\prime}},e^{i\beta^{\prime}},$ $e^{i\gamma^{\prime}})^{T}/2$
where $0\leq\alpha^{\prime},\beta^{\prime},\gamma^{\prime}<2\pi$. For
convenience, we will use an \emph{enphased} variant of $v$. Multiplying
through by the phase factor $e^{-i\alpha^{\prime}/2}$ and defining $%
\alpha=\alpha^{\prime}/2\in\left[0,\pi\right]$, $\beta=\beta^{\prime}-%
\alpha^{\prime}/2\in[0,2\pi]$, and similarly for $\gamma$, we consider the
parametrization $v=(e^{-i\alpha},e^{i\alpha},e^{i\beta},$ $e^{i\gamma})^{T}/2
$ instead. The conditions for $v(\alpha,\beta,\gamma)$ to be MU to the
columns of $F_{4}(x)$ lead to four equations, 
\begin{eqnarray}
\left\vert \cos\alpha\pm\zeta_{+}\right\vert & = & 1\,,  \label{zetaplus} \\
\left\vert \sin\alpha\pm e^{-ix}\zeta_{-}\right\vert & = & 1\,,
\label{zetaminus}
\end{eqnarray}
where complex numbers $\zeta_{\pm}=(e^{i\beta}\pm e^{i\gamma})/2$ have been
introduced. We will now construct all solutions of these equations as a
function of the value of $x$. We treat the cases (i) $\alpha=0$, (ii) $%
\alpha=\pi/2$, and (iii) $\alpha\neq0,\alpha \neq \pi/2$ separately since
the Eqs. (\ref{zetaplus}) and (\ref{zetaminus}) take different forms for
these values.

\subsection*{(i): $\protect\alpha=0$}

Eqs. (\ref{zetaminus}) simplify to the pair $\left\vert \pm
e^{-ix}\zeta_{-}\right\vert =1\,,$ which only hold simultaneously if $%
|\zeta_{-}|=1$ or $e^{i\gamma}=-e^{i\beta}$, implying that $\zeta_{+}=0$ so
that Eqs. (\ref{zetaplus}) are satisfied automatically. Thus, solutions
exist for any value of $x$ whenever $\beta=\gamma + \pi \rm{\,\, mod \, }{2\pi}$, 
and the resulting vectors can be
written as $v(\beta)=\left(1,1,e^{i\beta},-e^{i\beta}\right)^{T}/2$, with $%
\beta\in[0,2\pi]$. It will be convenient to divide this family of states
into two sets,%
\begin{equation}
h_{1}(y)=\frac{1}{2}\left(%
\begin{array}{c}
1 \\ 
1 \\ 
e^{iy} \\ 
-e^{iy}%
\end{array}%
\right),\, h_{2}(y^{\prime}) =\frac{1}{2}\left(%
\begin{array}{c}
1 \\ 
1 \\ 
-e^{iy^{\prime}} \\ 
e^{iy^{\prime}}%
\end{array}%
\right)\,,\qquad0\leq y,y^{\prime}<\pi,  \label{h1 and h2}
\end{equation}
introducing $y=\beta$ and $y^{\prime}=\pi+\beta$.

\subsection*{(ii): $\protect\alpha=\protect\pi/2$}

Eqs. (\ref{zetaplus}) and (\ref{zetaminus}) now reverse their roles: the
conditions $|\pm\zeta_{+}|=1$ require $e^{i\gamma}=e^{i\beta}$, with (\ref%
{zetaplus}) being satisfied since $|\zeta_{-}|=1$ follows immediately.
Hence, there is another one-parameter family of mutually unbiased vectors
for all values of $x$ if $\beta=\gamma$. This family can be written as $%
v(\varphi)=\left(1,-1,e^{i\varphi},e^{i\varphi}\right)^{T}/2$, $%
\varphi\in[0,2\pi]$, after dephasing and absorbing a factor of $i$ in the
definition of the phase, $\varphi=\pi/2+\beta$. Again, we express these
solutions as a set of pairs, 
\begin{equation}
h_{3}(z)=\frac{1}{2}\left(%
\begin{array}{c}
1 \\ 
-1 \\ 
e^{iz} \\ 
e^{iz}%
\end{array}%
\right),\, h_{4}(z^{\prime})=\frac{1}{2}\left(%
\begin{array}{c}
1 \\ 
-1 \\ 
-e^{iz^{\prime}} \\ 
-e^{iz^{\prime}}%
\end{array}%
\right)\,,\qquad0\leq z,z^{\prime}<\pi,  \label{h3 and h4}
\end{equation}
where $z=\varphi$ and $z^{\prime}=\pi+\varphi$.

\subsection*{(iii) $\protect\alpha\neq0,\protect\alpha \neq \protect\pi/2$}

\begin{figure}[ht]
\begin{center}
\includegraphics[width=0.4\textwidth]{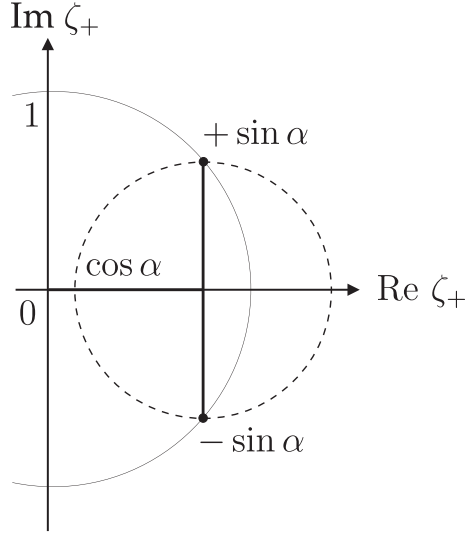} 
\end{center}
\caption{Plot of Eqs. (\protect\ref{zetaplus}) in the $\protect\zeta_{+}$%
-plane implying that, for $\cos \protect\alpha \neq 0$, their solutions are
given by $\protect\zeta_{+}=\mathrm{\pm {\it i}\sin\protect\alpha}$.}
\label{fourplane}
\end{figure}

A plot in the complex plane (see Fig. \ref{fourplane}) reveals that one must
have $\zeta_{+}=\mathrm{\pm {\it i}\sin\alpha}$ if Eqs. (\ref{zetaplus}) are to
hold with $\cos\alpha\neq0$. Thus, the real part of $\zeta_{+}$ vanishes, 
\begin{equation}
\cos\beta+\cos\gamma=0  \label{betagamma}
\end{equation}
with $\gamma=\pi-\beta \rm{\,\, mod \, }{2\pi}$
being the only acceptable solution: the other solution, $\gamma=\pi+\beta\rm{\,\, mod \, }{2\pi}$ leads to $0=\zeta_{+}=%
\mathrm{\pm i\sin\alpha}$, producing a contradiction since $\alpha\neq0$.
Thus, using $\gamma=\pi-\beta \rm{\,\, mod \, }{2\pi}$, we obtain $\zeta_{+}=i\sin\beta$ find the following relation
between $\alpha$ and $\beta$: 
\begin{equation}
\pm\sin\alpha=\sin\beta\,.  \label{final cond 1}
\end{equation}

Similarly, Eqs. (\ref{zetaminus}) for $\sin\alpha\neq0$ imply that $%
e^{-ix}\zeta_{-}=\pm i\cos\alpha$. Using $\gamma=\pi-\beta \rm{\,\, mod \, }{2\pi}$ in the definition of $\zeta_{-}$, we
find $\zeta_{-}=\cos\beta$, so that

\begin{equation}
i(\pm\cos\alpha+\sin x\cos\beta)=\cos x\cos\beta.
\label{nearly final cond 2}
\end{equation}
The right-hand-side of this equation only vanishes if $x=\pi/2$: both $%
\beta=\pi/2$ and $\beta=3\pi/2$ would, according to (\ref{final cond 1}),
require $\alpha=\pi/2$ which we currently exclude. Therefore, solutions to
Eqs. (\ref{zetaplus},\ref{zetaminus}) with $\alpha\neq0$ or $\alpha\neq\pi/2$
only exist for $x=\pi/2$ if a second relation between $\alpha$ and $\beta$
holds,

\begin{equation}
\pm\cos\alpha=\cos\beta\,.  \label{final cond 2}
\end{equation}

The form of the additional MU vectors is determined by Eqs. (\ref{final cond 1}) and (\ref{final cond 2}) which have four solutions. First, for $%
\beta=\alpha$ we obtain MU vectors of the form $(e^{-i\alpha},e^{i%
\alpha},e^{i\alpha},$ $-e^{ia})^{T}/2$ or $\left(1,e^{2i\alpha},e^{2i%
\alpha},-1\right)^{T}/2$ after dephasing. Splitting this family into two
subsets as before, we find 
\begin{equation}
k_{1}=\frac{1}{2}\left(%
\begin{array}{c}
1 \\ 
e^{it} \\ 
e^{it} \\ 
-1%
\end{array}%
\right)\,,\; k_{2}=\frac{1}{2}\left(%
\begin{array}{c}
1 \\ 
-e^{it^{\prime}} \\ 
-e^{it^{\prime}} \\ 
-1%
\end{array}%
\right)\,,\quad0\leq t,t^{\prime}<\pi\,.  \label{k1 k2}
\end{equation}
Similarly, the choice $\beta=\pi+\alpha \rm{\,\, mod \, }{2\pi}$ leads to two sets of dephased MU vectors, 
\begin{equation}
k_{3}=\frac{1}{2}\left(%
\begin{array}{c}
1 \\ 
e^{iu} \\ 
-e^{iu} \\ 
-1%
\end{array}%
\right)\,,\; k_{4}=\frac{1}{2}\left(%
\begin{array}{c}
1 \\ 
-e^{iu^{\prime}} \\ 
e^{iu^{\prime}} \\ 
-1%
\end{array}%
\right)\,,\quad0\leq u,u^{\prime}<\pi\,.  \label{k3 k4}
\end{equation}
Next, when proceeding in an entirely analogous manner for the remaining two
choices $\beta=\pi-\alpha \rm{\,\, mod \, }{2\pi}$ and $\beta=2\pi-\alpha \rm{\,\, mod \, }{2\pi}$%
, we obtain the following four families of dephased vectors MU to $%
F_{4}(\pi/2)$, 
\begin{equation}
j_{1}=\frac{1}{2}\left(%
\begin{array}{c}
1 \\ 
e^{ir} \\ 
-1 \\ 
e^{ir}%
\end{array}%
\right)\,,\; j_{2}=\frac{1}{2}\left(%
\begin{array}{c}
1 \\ 
-e^{ir^{\prime}} \\ 
-1 \\ 
-e^{ir^{\prime}}%
\end{array}%
\right)\,,\; j_{3}=\frac{1}{2}\left(%
\begin{array}{c}
1 \\ 
e^{is} \\ 
1 \\ 
-e^{is}%
\end{array}%
\right)\,,\; j_{4}=\frac{1}{2}\left(%
\begin{array}{c}
1 \\ 
-e^{is^{\prime}} \\ 
1 \\ 
e^{is^{\prime}}%
\end{array}%
\right),  \label{j family}
\end{equation}
with $0\leq r,r^{\prime},s,s^{\prime}<\pi$.

\subsection{Forming MU bases}

Knowing all vectors that are MU to both the identity and $F_{4}$, we now
determine those combinations which form other bases.

\subsection*{Triples of MU bases in $\mathbb{C}^4$}

To begin, consider the MU vectors $h_{1},\ldots,h_{4},$ in Eqs. (\ref{h1 and h2},\ref{h3 and h4}) which exist for all values of $x\in[0,\pi]$.
Calculating their inner products, one finds that they only form an \emph{%
orthonormal} basis of $\mathbb{\mathbb{C}}^{4}$ if $y=y^{\prime}$ and $%
z=z^{\prime}$. Thus, for each value of $x$, the pair $\left\{
I,F_{4}(x)\right\} $ may be complemented by a third MU basis taken from the
two-parameter family 
\begin{equation}  \label{H(y,z)}
H_{4}(y,z)=\frac{1}{2}\left(%
\begin{array}{cccc}
1 & 1 & 1 & 1 \\ 
1 & 1 & -1 & -1 \\ 
-e^{iy} & e^{iy} & e^{iz} & -e^{iz} \\ 
e^{iy} & -e^{iy} & e^{iz} & -e^{iz}%
\end{array}%
\right)\,.
\end{equation}
In other words, there is a \emph{three parameter-family of triplets}
of MU bases $\left\{ I,F_{4}(x),H_{4}(y,z)\right\} $ in
dimension $d=4$. This family was not known before. 

If $x=\pi/2$, additional MU vectors $j_{1},\ldots,j_{4}$, and $%
k_{1},\ldots,k_{4}$, have been identified, cf. Eqs. (\ref{k1 k2}-\ref{j family}). Calculating the scalar products within each group, one sees that
two further orthonormal two-parameter bases emerge, 
\begin{eqnarray}  \label{J4}
J_{4}(r,s) & = & \frac{1}{2}\left(%
\begin{array}{cccc}
1 & 1 & 1 & 1 \\ 
e^{ir} & -e^{ir} & e^{is} & -e^{is} \\ 
-1 & -1 & 1 & 1 \\ 
e^{ir} & -e^{ir} & -e^{is} & e^{is}%
\end{array}%
\right)\,, \\
K_{4}(t,u) & = & \frac{1}{2}\left(%
\begin{array}{cccc}
1 & 1 & 1 & 1 \\ 
e^{it} & -e^{it} & e^{iu} & -e^{iu} \\ 
e^{it} & -e^{it} & -e^{iu} & e^{iu} \\ 
-1 & -1 & 1 & 1%
\end{array}%
\right)\,,  \label{K4}
\end{eqnarray}
if the conditions $r=r^{\prime}$, $s=s^{\prime}$, and $t=t^{\prime}$, $%
u=u^{\prime}$, respectively, are satisfied. No other combinations of the MU
vectors can form inequivalent bases so that the matrices in Eqs. (\ref%
{H(y,z)}-\ref{K4}) represent all possible choices of a MU basis. Permuting
appropriate rows and columns of the matrices $J_{4}$ and $K_{4}$ transforms
them into $H_{4}$; thus, the triples $\left\{
I,F_{4}(\pi/2),J_{4}(r,s)\right\} $ and $\left\{
I,F_{4}(\pi/2),K_{4}(t,u)\right\} $ are equivalent to $\left\{
I,F_{4}(\pi/2),H_{4}(y,z)\right\} $.

\subsection*{Quadruples and quintuples of MU bases in $\mathbb{C}^4$}

Let us begin by noting that sets of four MU bases cannot exist away from $%
x=\pi/2$. No two matrices $H_{4}(y,z)$ and $H_{4}(y^{\prime},z^{\prime})$
are MU since 
\begin{equation}
\left\vert h_{1}^{\dagger}(y)h_{1}(y^{\prime})\right\vert =\left\vert
h_{1}^{\dagger}(y)h_{2}(y^{\prime})\right\vert =\frac{1}{2}  \label{H4 test}
\end{equation}
only hold if 
\begin{equation}
\left\vert 1\pm e^{i(y-y^{\prime})}\right\vert =1\,;
\label{H4 contradiction}
\end{equation}
however, these equations have no solution for any values of $y$ and $%
y^{\prime}$. A similar argument shows that there are no values of $z$ and $%
z^{\prime}$ such that the matrices $H_{4}(y,z)$ and $H_{4}(y,z^{\prime})$
are MU.

We now show that for $x=\pi/2$ the bases $H_{4}(y,z),J_{4}(r,s)$ and $%
K_{4}(t,u)$ give rise to four and five MU bases if the free parameters are
chosen appropriately. An argument similar to the one just presented shows
that no two bases within either the family $J_{4}(r,s)$ or $K_{4}(t,u)$ are
MU. Thus, any quadruple of MU bases must contain bases from different
families.

The inner products $\left\vert h_{1}^{\dagger}(y)j_{1}(r)\right\vert
,\left\vert h_{1}^{\dagger}(y)j_{2}(r)\right\vert ,\left\vert
h_{2}^{\dagger}(y)j_{1}(r)\right\vert $ and $\left\vert
h_{2}^{\dagger}(y)j_{2}(r)\right\vert $ have modulus $1/2$ if there are
values for $y$ and $r$ such that the equations 
\begin{eqnarray}  \label{d=4 constraints}
\left\vert 1+e^{ir} \pm (e^{-iy} + e^{i(r-y)})\right\vert  =  2& & \\
\left\vert 1-e^{ir} \pm (e^{-iy} - e^{i(r-y)})\right\vert  =  2& & 
\end{eqnarray}
hold simultaneously. Upon introducing a factor of $e^{-ir/2}$, Eqs. (\ref%
{d=4 constraints}) are equivalent to the constraints 
\begin{equation}
\left\vert \cos\frac{r}{2}\right\vert =  \frac{1}{\left|1\pm
e^{-iy}\right|}=\left\vert \sin\frac{r}{2}\right\vert \,.
\label{quadruple conditions}
\end{equation}
Consequently, one must have $r=\pi/2$ , and thus $e^{-iy}=\pm i$ or $y=\pi/2$
since $0\leq r,y<\pi$. An entirely analogous argument restricts the values
of $s$ and $z$: checking the inner products $\left\vert
h_{1}(r)^{\dagger}j_{3}(z)\right\vert ,\left\vert
h_{1}(r)^{\dagger}j_{4}(z)\right\vert $ etc. tells us that the matrices $%
H_{4}(y,z)$ and $J_{4}(r,s)$ are mutually unbiased only if $y=z=r=s=\pi/2$.
We also find that the pairs $\{ J_{4}(r,s),$ $K_{4}(t,u)\} $ and $\left\{
K_{4}(t,u),H_{4}(x,y)\right\} $ are MU only when all six parameters take the
value $\pi/2.$

We are now in the position to list all possible sets of MU bases in $\mathbb{%
C}^{4}$ beyond $\{I,F_{4}(x)\}$, 
\begin{eqnarray}
\{I,F_{4}(x),H_{4}(y,z)\}\,,  \nonumber \\
\{I,F_{4}(\pi/2),H_{4}(\pi/2,\pi/2),J_{4}(\pi/2,\pi/2)\}\,,  \label{d=4 list}
\\
\{I,F_{4}(\pi/2),H_{4}(\pi/2,\pi/2),J_{4}(\pi/2,\pi/2),K_{4}(\pi/2,\pi/2)\}%
\,.  \nonumber
\end{eqnarray}
There is one three-parameter family of \emph{triples} consisting of the
one-parameter Fourier family $F_{4}(x)$ combined with two-parameter set $%
H_{4}(y,z)$; neither $J_{4}(r,s)$ nor $K_{4}(t,u)$ give rise to other
triples since each of these sets of Hadamard matrices is equivalent to $%
\{I,F(\pi/2),H_4(y,z)\}$. The three-dimensional set (\ref{d=4 list}) of MU bases
in dimension $d=4$ may be visualized as a cuboid defined by $%
0\leq x,y<\pi $ and $0\leq z<\pi /2$. The reduction in the parameter range
of $z$ is due to the equivalence $\{I,F_{4}(x),H_{4}(y,z)\}\,\sim
\{I,F_{4}(\pi -x),H_{4}(\pi -y,\pi -z)\}\,$\ which follows from an overall
complex conjugation. Each of the points in the cuboid corresponds to one
triple while both the quadruple and the quintuple are located at the point, $%
x=y=z=\pi /2$.

Only one set of \emph{four} MU bases exists, $\{I,F_{4}(\pi/2),H_{4}(\pi/2,%
\pi/2),J_{4}(\pi/2,\pi/2)\}$, since the other two candidates obtained by
combining $K_{4}(\pi/2,\pi/2)$ with either $J_{4}(\pi/2,\pi/2)$ or $%
H_{4}(\pi/2,\pi/2)$ are permutations of this quadruple. Finally, there is a
unique way to a construct \emph{five} MU bases which is easily seen to be
equivalent to the standard construction of a complete set of MU bases in
dimension four.

\section{Dimension $d=5$}

As in dimensions two and three, there is a unique choice of a $(5\times 5)$
dephased complex Hadamard matrix \cite{haagerup96}, 
\begin{equation}  \label{f5}
F_{5}=\frac{1}{\sqrt{5}}\left( 
\begin{array}{ccccc}
1 & 1 & 1 & 1 & 1 \\ 
1 & \omega & \omega ^{2} & \omega ^{3} & \omega ^{4} \\ 
1 & \omega ^{2} & \omega ^{4} & \omega & \omega ^{3} \\ 
1 & \omega ^{3} & \omega & \omega ^{4} & \omega ^{2} \\ 
1 & \omega ^{4} & \omega ^{3} & \omega ^{2} & \omega%
\end{array}
\right) ,
\end{equation}
equal to the discrete $(5\times 5)$ Fourier matrix, with $\omega =\exp (2\pi
i/5)$ denoting a fifth root of unity. The uniqueness of $F_{5}$ is obtained by an analytical method that is far from elementary \cite{haagerup96}.

We have not found an elementary method to obtain a list of all vectors which
are MU to the Fourier matrix $F_5$. Instead, we will rely on earlier work
\cite{brierley+09} where those vectors have been constructed 
\emph{analytically} by means of a computer program.

\subsection{Constructing vectors MU to $F_5$}

The vector $v=(1,e^{i\alpha _{1}},\ldots ,e^{i\alpha _{4}})/\sqrt{5}\in 
\mathbb{C}^{5}$ is MU to $F_{5}$ if it satisfies the conditions 
\begin{equation}
\left\vert \sum_{j=0}^{4}\omega ^{jk}e^{i\alpha _{j}}\right\vert =\sqrt{5}%
\,,\quad k=0\ldots 4\,,  \label{d=5 constraints}
\end{equation}%
defining $\alpha _{0}\equiv 0$. According to \cite{brierley+09}, the
solutions of these equations give rise to 20 vectors which can be arranged
in four MU bases, 
\begin{eqnarray}
H_{5}^{(1)} &=&\frac{1}{\sqrt{5}}\left( 
\begin{array}{ccccc}
1 & 1 & 1 & 1 & 1 \\ 
\omega  & \omega ^{2} & \omega ^{3} & \omega ^{4} & 1 \\ 
\omega ^{4} & \omega  & \omega ^{3} & 1 & \omega ^{2} \\ 
\omega ^{4} & \omega ^{2} & 1 & \omega ^{3} & \omega  \\ 
\omega  & 1 & \omega ^{4} & \omega ^{3} & \omega ^{2}%
\end{array}%
\right) ,\;H_{5}^{(2)}=\frac{1}{\sqrt{5}}\left( 
\begin{array}{ccccc}
1 & 1 & 1 & 1 & 1 \\ 
\omega ^{2} & \omega ^{3} & \omega ^{4} & 1 & \omega  \\ 
\omega ^{3} & 1 & \omega ^{2} & \omega ^{4} & \omega  \\ 
\omega ^{3} & \omega  & \omega ^{4} & \omega ^{2} & 1 \\ 
\omega ^{2} & \omega  & 1 & \omega ^{4} & \omega ^{3}%
\end{array}%
\right) ,  \nonumber \\
H_{5}^{(3)} &=&\frac{1}{\sqrt{5}}\left( 
\begin{array}{ccccc}
1 & 1 & 1 & 1 & 1 \\ 
\omega ^{3} & \omega ^{4} & 1 & \omega  & \omega ^{2} \\ 
\omega ^{2} & \omega ^{4} & \omega  & \omega ^{3} & 1 \\ 
\omega ^{2} & 1 & \omega ^{3} & \omega  & \omega ^{4} \\ 
\omega ^{3} & \omega ^{2} & \omega  & 1 & \omega ^{4}%
\end{array}%
\right) ,\;H_{5}^{(4)}=\frac{1}{\sqrt{5}}\left( 
\begin{array}{ccccc}
1 & 1 & 1 & 1 & 1 \\ 
\omega ^{4} & 1 & \omega  & \omega ^{2} & \omega ^{3} \\ 
\omega  & \omega ^{3} & 1 & \omega ^{2} & \omega ^{4} \\ 
\omega  & \omega ^{4} & \omega ^{2} & 1 & \omega ^{3} \\ 
\omega ^{4} & \omega ^{3} & \omega ^{2} & \omega  & 1%
\end{array}%
\right) .  \label{d=5 complete set}
\end{eqnarray}%
To obtain this result, Eqs. (\ref{d=5 constraints}) have been expressed as a
set of coupled quadratic polynomials in eight real variables. Using an
implementation \cite{salsa} of Buchberger's algorithm \cite{buchberger65,buchberger98} on the computer program Maple \cite{Maple}, a Gr\"{o}bner basis of these equations has been constructed which leads to the 20
vectors given by the columns of the four Hadamard matrices above. It is
important to note that no other solutions of Eqs. (\ref{d=5 constraints})
exist, a result which does \emph{not} follow from the known methods to
construct a complete set of six MU bases in $\mathbb{C}^{5}$.

Each of the four matrices in (\ref{d=5 complete set}) is related to the
Fourier matrix in a remarkably simple manner. In analogy to the unitary
diagonal matrix used in Eq. (\ref{powers of D}), define a diagonal unitary
matrix 
\begin{equation}
D=\mbox{diag}(1,\omega ,\omega ^{4},\omega ^{4},\omega )\,,
\label{diagonalD}
\end{equation}%
with entries given by the first column of $H_{5}^{(1)}$ and you find that 
\begin{equation}
H_{5}^{(k)}=D^{k}F_{5}\,,\quad k=1,\ldots ,4\,.  \label{enphased fourier}
\end{equation}%
Using this observation, we can express the unique complete set of six MU
bases for dimension $d=5$ as follows 
\begin{equation}
\{I,F_{5},H_{5}^{(1)},\ldots ,H_{5}^{(4)}\}\equiv
\{I,F_{5},DF_{5},D^{2}F_{5},D^{3}F_{5},D^{4}F_{5}\}\,,  \label{complete d=5}
\end{equation}%
which will be useful later on.

Next, we proceed to classify all smaller sets of MU bases of $\mathbb{C}^5$
by combining subsets of the four Hadamard matrices $H_5^{(k)}$ in (\ref{d=5 complete set}) with the pair $\{I,F_5\}$. For clarity, we now list the set
of inequivalent classes which we will obtain. In addition to the pair $%
\{I,F_5\}$ and the complete set given in (\ref{complete d=5}) there are 
\emph{two} inequivalent triples as well as \emph{one} quadruple and \emph{one}
quintuple: 
\begin{eqnarray}
\{I,F_{5},H_{5}^{(1)}\} \, , \{I,F_{5},H_{5}^{(2)}\}\,,  \nonumber \\
\{I, F_5, H_5^{(1)}, H_5^{(2)}\} \,,  \label{d=5 list} \\
\{I, F_5, H_5^{(1)}, H_5^{(2)},H_5^{(3)}\} \,.  \nonumber
\end{eqnarray}

\subsection*{Triples of MU bases in $\mathbb{C}^5$}

Select one of the four matrices given in (\ref{d=5 complete set}) and adjoin
it to the pair $\{I,F_5\}$. You obtain four triples of MU bases with two immediate equivalences, namely, 
\begin{equation}  \label{triples in C5 a}
\{I,F_{5},H_{5}^{(1)}\} \equiv \{I,F_{5},DF_{5}\} \sim \{I,F_{5},
D^{4}F_{5}\} \equiv \{I,F_{5},H_{5}^{(4)}\}
\end{equation}
on the one hand, and 
\begin{equation}  \label{triples in C5 b}
\{I,F_{5},H_{5}^{(2)}\} \equiv \{I,F_{5},D^2F_{5}\} \sim
\{I,F_{5},D^{3}F_{5}\} \equiv \{I,F_{5},H_{5}^{(3)}\}
\end{equation}
on the other. The equivalence (\ref{triples in C5 a}) follows from
multiplying the set $\{I,F_{5},DF_{5}\}$ with $D^4$ from the left, rephasing
the first basis with $D$ from the right, using $D^5=I$ and swapping the last
two matrices. A similar argument establishes the equivalence (\ref{triples in C5 b}), using $D^3$ instead of $D^4$.

Thus, it remains to check whether the triples $\mathcal{T}^{(1)} \equiv
\{I,F_{5},H_{5}^{(1)}\}$ and $\mathcal{T}^{(2)} \equiv \{I,F_{5},$ $%
H_{5}^{(2)}\}$ are equivalent to each other. It turns out that these two
triples are, in fact, \emph{inequivalent}. More explicitly, this means that
no unitary matrix $U$ and no monomial matrices $M_0,M_1$ and $M_2$ can be
found which would map $\mathcal{T}^{(1)} $ into $\mathcal{T}^{(2)}$
according to 
\begin{equation}  \label{inequivalence for triples 1}
\{I,F_{5},H_{5}^{(1)}\} \rightarrow \{U I M_{0},U F_{5} M_{1},U H_{5}^{(1)}
M_{2}\} \, .
\end{equation}
A proof of this statement is given in Appendix B.

\subsection*{Quadruples of MU bases in $\mathbb{C}^5$}

There are six possibilities to form quadruples by selecting two of the four
matrices in Eq. (\ref{d=5 complete set}) and adding them to the pair $%
\{I,F_5\}$. Recalling that $H_5^{(k)} = D^k F_5$, we identify the following
equivalences which relate three quadruples each, 
\begin{equation}  \label{d=5equivquintquad}
\{I, F_5, D F_5, D^2 F_5\} \sim \{I,F_5, D^3 F_5, D^4 F_5\} \sim \{I, F_5, D
F_5, D^4 F_5\} \, , 
\end{equation}
and 
\begin{equation}  \label{d=5equivquintquad2}
\{I,F_5, D F_5, D^3 F_5\} \sim \{I, F_5, D^2 F_5, D^4 F_5\} \sim \{I, F_5,
D^2 F_5, D^3 F_5\} \, . 
\end{equation}
To show the first equivalence in Eq. (\ref{d=5equivquintquad}), for example,
multiply its left-hand-side with $D^3$ from the left, use the identity $D^5 = 1$ and
rearrange the bases appropriately. The other equivalences follow from
analogous arguments. Thus, there are at most two inequivalent sets of four
MU bases in $\mathbb{C}^5$, with representatives $\{I, F_5, H_5^{(1)},
H_5^{(2)}\}$ and $\{I, F_5, H_5^{(1)}, H_5^{(3)}\}$, say.

Interestingly, these two classes of MU bases are \emph{equivalent} to each
other leaving us with a single equivalence class of quadruples in dimension
five, with representative $\{I,F_{5},H_{5}^{(1)},H_{5}^{(2)}\}$, say. To
show this equivalence, we multiply the first quadruple with the adjoint of $%
F_{5}$ from the left 
\begin{equation}
F_{5}^{\dagger }\{I,F_{5},H_{5}^{(1)},H_{5}^{(2)}\}\sim
\{I,F_{5},F_{5}^{\dagger }H_{5}^{(1)},F_{5}^{\dagger }H_{5}^{(2)}\}\,.
\label{d=5last equivalence}
\end{equation}%
using the identity $F^{\dagger }=FP$, with some permutation matrix $P$, and 
swapping the first two bases. The
action of $F_{5}^{\dagger }$ on the other two elements is surprisingly
simple: the Hadamard matrix $H_{5}^{(1)}$ is mapped to itself, 
\begin{equation}
F_{5}^{\dagger }H_{5}^{(1)}=H_{5}^{(1)}M\,,  \label{fdaggeraction2}
\end{equation}%
up to a monomial matrix $M$, while $H_{5}^{(2)}$ is sent to $H_{5}^{(3)}$, 
\begin{equation}
F_{5}^{\dagger }H_{5}^{(2)}=H_{5}^{(3)}M^{\prime }\,,  \label{fdaggeraction1}
\end{equation}%
again up to some monomial matrix $M^\prime$. Both relations simply follow from working
out the product on the left and factoring the result, e.g. 
\begin{equation}
F_{5}^{\dagger }H_{5}^{(2)}=\frac{1}{\sqrt{5}}\left( 
\begin{array}{rrrrr}
s(1) & s(2) & s(3) & s(4) & s(5) \\ 
\omega ^{3}s(1) & \omega ^{2}s(2) & \omega s(3) & s(4) & \omega ^{4}s(5) \\ 
\omega ^{2}s(1) & s(2) & \omega ^{3}s(3) & \omega s(4) & \omega ^{4}s(5) \\ 
\omega ^{2}s(1) & \omega ^{4}s(2) & \omega s(3) & \omega ^{3}s(4) & s(5) \\ 
\omega ^{3}s(1) & \omega ^{4}s(2) & s(3) & \omega s(4) & \omega^{2} s(5)%
\end{array}%
\right) =H_{5}^{(3)}D^{(2)}\,P,
\end{equation}%
where the $k^{th}$ entry of the diagonal matrix $D^{(2)}$ is given by the
sum of the $k^{th}$ column of $H_{5}^{(2)}$, denoted by $s(k)=%
\sum_{i}H_{ik}^{(2)}$, and $P$ permutes the columns. Using these identities
in Eq. (\ref{d=5last equivalence}) we find that 
\begin{equation}
\{I,F_{5},H_{5}^{(1)},H_{5}^{(2)}\}\sim \{I,F_{5},H_{5}^{(1)},H_{5}^{(3)}\}\,
\end{equation}%
the two quadruples are equivalent.

\subsection*{Quintuples of MU bases in $\mathbb{C}^5$}

Four sets of MU bases can be obtained by adding any three of the four
matrices in Eq. (\ref{d=5 complete set}) to the pair $\{I,F_5\}$. It is not
difficult to show that the four resulting sets of quintuples are equivalent
to each other. Thus, there is effectively only one possibility to choose
five MU bases in $\mathbb{C}^5$, with representative $\{I, F_5, H_5^{(1)},
H_5^{(2)},H_5^{(3)}\}$.

Let us show now that this representative, which has been obtained by leaving
out $H_5^{(4)}$, is equivalent to the set $\{I, F_5, H_5^{(1)},
H_5^{(2)},H_5^{(4)}\}$, for example. Indeed, the equivalence 
\begin{equation}  \label{d=5equivquint}
\{I,F_5, D F_5, D^2 F_5, D^3 F_5\} \sim \{I, F_5, D F_5, D^2 F_5, D^4 F_5\}
\, , 
\end{equation}
follows immediately from multiplying the second set by $D$ from the left and
using $D^5 = I$, 
\begin{equation}  \label{d=5equivquint2}
\{I, F_5, D F_5, D^3 F_5, D^4 F_5\} \sim \{I, D F_5, D^2 F_5, D^3 F_5, F_5\}
\, . 
\end{equation}
Reordering the set of five matrices on the right reveals the desired
equivalence with the quintuple $\{I, F_5, H_5^{(1)}, H_5^{(2)},H_5^{(3)}\}$.
Effectively, the four matrices different from $I$ undergo a cyclic shift
under multiplication with $D$, and the remaining equivalences follow from
shifts induced by $D^2$ and $D^3$, respectively.

\section{Summary and Discussion\label{concl}}

We have constructed all inequivalent sets of mutually unbiased bases in
dimension two to five. Our approach is based on the fact that all complex
Hadamard matrices are known in these dimensions. For dimensions up to $d=4$,
elementary arguments suffice to classify the existing sets of MU bases while
dimension five requires some analytic results which have been found earlier
using algebraic computer software.

\begin{table}[h!]
\begin{center}
\begin{tabular}{rccccccc}
\hline
$d$ &  & $2$ & $3$ & $4 $ & $5$ &  & $6$ \\ \hline
pairs &  & 1 & 1 & $\infty^1$ & 1 &  & $\geq \infty^3$ \\ 
triples &  & 1 & 1 & $\infty^3$ & 2 &  & $\geq \infty^2$ \\ 
quadruples &  & - & 1 & 1 & 1 &  & ? \\ 
quintuples &  & - & - & 1 & 1 &  & ? \\ 
sextuples &  & - & - & - & 1 &  & ? \\ \hline
\end{tabular}%
\end{center}
\caption{The number of inequivalent MU bases for dimensions two to six where 
$\infty^k$ denotes a $k$-parameter set; see text for details.}
\label{MU classes}
\end{table}

The first four columns of Table \ref{MU classes} summarize the results
obtained in this paper. All \emph{pairs} of MU bases in dimensions two to
five are listed in the first row, effectively reflecting the known
classification of inequivalent Hadamard matrices; a continuous
(one-parameter) set of inequivalent MU pairs only exists in dimension four.

The main results concern \emph{triples} of MU bases in
dimension four where we find a \emph{three-parameter family} and in
dimension five where we obtain \emph{two} inequivalent triples. 

Finally, we have shown that there is only one class of both MU \emph{%
quadruples} and MU \emph{quintuples} in dimensions four and five. In all 
dimensions considered, there is a unique $d$-tuple which can be extended 
to a complete set of $(d+1)$ MU bases using a construction presented in \cite{weiner09}.

The last column of Table \ref{MU classes} contrasts these results with
dimension six where the classification of all complex Hadamard 
matrices is not known to be complete. The first entry shows that 
there is a three-parameter family of pairs of MU
bases \cite{karlsson10} (it has been conjectured that the parameter
space has, in fact, four dimensions \cite{skinner+09}). Furthermore, it is possible to construct a two-parameter family of
triples \cite{szollosi08} using an idea taken from \cite{zauner99}. There 
is strong numerical \cite{butterley+07, brierley+08} and analytical 
\cite{brierley+09,Jaming+09} evidence to suggest that triples are the largest
sets of MU bases in dimension six but this problem remains open.

The notion of equivalence used in this paper (see Appendix \ref{app A}) is
mathematical in nature; it captures all possible operations that leave invariant the
conditions (\ref{MUB conditions}) for two bases to be mutually unbiased.
Motivated by experiments, there is a finer equivalence of complete sets of
MU bases based on the entanglement structure of the states contained in each
basis \cite{lawrence+02, romero+05}. For dimensions that are a power of two,
a complete set of MU bases can be realized using Pauli operators acting on
each two-dimensional subsystem. Two sets of MU bases are then called
equivalent when they can be factored into the same number of subsystems. For 
$d=2,4$ this notion of equivalence also leads to a unique set of $(d+1)$ MU
bases. However, for $d=8,16, \ldots$ complete sets of MU bases can have
different entanglement structures even though they are equivalent up to an
overall unitary transformation \cite{lawrence+02, romero+05}.

The traditional approach to find \emph{complete} sets of MU bases in
prime-power dimensions via the Heisenberg-Weyl group or by using finite fields is
constructive and, therefore, does not exclude the existence of other
inequivalent complete sets. The present approach is, in contrast, \emph{%
exhaustive}: we are able to affirm that the known complete sets for $2\leq d
\leq5$ are unique (up to equivalence). Their uniqueness has been shown
earlier for $d \leq 4$ \cite{ericsson04} while \cite{boykin+05} contains a proof 
for $2 \leq d \leq 5$ in a Lie algebraic setting. We find it appealing that it is 
possible to prove the uniqueness of complete sets of MU bases in low
dimensions by \emph{elementary} methods.

\section*{Acknowledgements}
\noindent 
The authors would like to thank M. Kibler for pointing out the complex conjuation equivalence relation given in Eq. (\ref{conjugationequiv}). 
SB and SW gratefully acknowledge financial support through QNET, a network
funded by the EPSRC, and IB has been supported by the Swedish Research Council.

\appendix
\section{Equivalent sets of MU bases \label{equivalences}}

\label{app A}

Many sets of MU bases are identical to each other. To simplify the
enumeration of all sets of MU bases we introduce \textit{equivalence classes}
and a \textit{standard form} of sets of MU bases. 

Each set of $(r+1)$ MU bases in $\mathbb{C}^{d}$ corresponds to a list of $%
(r+1)$ (with $r \leq d$) complex matrices $H_{\rho}$, $\rho=0,1,\ldots,r$ of
size $(d\times d)$. Two such lists $\{H_{0},H_{1},\ldots,H_{r}\}$ and $%
\{H_{0}^{\prime},H_{1}^{\prime},\ldots,H_{r}^{\prime}\}$ are \emph{equivalent%
} to each other, 
\begin{equation}
\{H_{0},H_{1},\ldots,H_{r}\}\sim\{H_{0}^{\prime},H_{1}^{\prime},%
\ldots,H_{r}^{\prime}\}  \label{MUequivalence}
\end{equation}
if they can be transformed into each other by a succession of the following
four transformations:

\begin{enumerate}
\item an \emph{overall unitary} transformation $U$ applied from the left, 
\begin{equation}
\{H_{0},H_{1},\ldots,H_{r}\}\to
U\{H_{0},H_{1},\ldots,H_{r}\}\equiv\{UH_{0},UH_{1},\ldots,UH_{r}\}\,,
\label{overallunitary}
\end{equation}
which leaves invariant the value of all scalar products;

\item $(r+1)$ \emph{diagonal unitary} transformations $D_{\rho}$ from the
right which attach phase factors to each column of the $(r+1)$ matrices, 
\begin{equation}  \label{diagunitaries}
\{H_{0},H_{1},\ldots,H_{r}\}\to\{H_{0}D_{0},H_{1}D_{1},\ldots,H_{r}D_{r}\}
\, ;
\end{equation}
these transformations exploit the fact that the overall phase of a quantum
state drops out of from the conditions of MU bases;

\item $(r+1)$ \emph{permutations} of the elements within each basis, 
\begin{equation}
\{H_{0},H_{1},\ldots,H_{r}\}\to\{H_{0}P_{0},H_{1}P_{1},\ldots,H_{r}P_{r}\}
\,,  \label{permutations}
\end{equation}
which amount to relabeling the elements within each basis by means of
unitary permutation matrices $P_{n}$ satisfying $PP^T=I$;

\item \emph{pairwise exchanges} of two bases, 
\begin{equation}
\{\ldots,H_{\rho},\ldots,H_{\rho^{\prime}},\ldots\}\to\{\ldots,H_{\rho^{%
\prime}},\ldots,H_{\rho},\ldots\}\,,  \label{swaps}
\end{equation}
which amounts to relabeling the bases.

\item an \emph{overall complex conjugation} 
\begin{equation}\label{conjugationequiv}
\{H_{0},H_{1},\ldots ,H_{r}\}\rightarrow \{{H}_{0}^*,{H}%
_{1}^*,\ldots ,{H}_{r}^*\}  \label{conjugation}
\end{equation}%
which leaves the values of all scalar products invariant.
\end{enumerate}

These equivalence relations allow us to \emph{dephase} a given set of MU
bases. The resulting \emph{standard form} $\{I,H_{1},\ldots,$ $H_{r}\}$ is
characterized by four properties: (i) the first basis is chosen to be the
standard basis of $\mathbb{C}^{d}$ described by $H_{0}\equiv I$, where $I$
is the $(d\times d)$ identity matrix; (ii) the remaining bases are described
by (complex) \emph{Hadamard} matrices: each of their matrix elements has
modulus $1/\sqrt{d}$; (iii) the components of the first column of the matrix 
$H_{1}$ are given by $1/\sqrt{d}$; (iv) the first row of each of the
Hadamard matrices $H_{1}$ to $H_{r}$ has entries $1/\sqrt{d}$ only.

\section{Inequivalent triples of MU bases in $\mathbb{C}^5$}

\label{app B}

We show that the two classes of triples of MU bases given by $\mathcal{T}%
^{(1)} \equiv \{I,F_{5},H_{5}^{(1)}\}$ and $\mathcal{T}^{(2)} \equiv
\{I,F_{5},H_{5}^{(2)}\}$ are \emph{inequivalent}. In a first step, we
explain that it is sufficient to search for equivalence transformations
generated by matrices of a special form. In a second step we show that a
contradiction arises if one assumes that the triples $\mathcal{T}^{(1)}$ and 
$\mathcal{T}^{(2)}$ are equivalent.

Let us begin with a general remark about the structure of equivalence
classes of sets of MU bases $\mathcal{M} = \{I,B_{1},\ldots ,B_{r}\}$ of $%
\mathbb{C}^d$ for all $r \in \{1,\ldots, d-1 \}$. For convenience, we assume that the first basis equals the
identity, i.e. the set is given in standard form. As explained in Appendix %
\ref{app A} all sets of MU bases equivalent to $\mathcal{M}$ are obtained as
follows, 
\begin{equation}
\mathcal{M}\rightarrow \mathcal{M}^\prime = \{UM_{0},UB_{1}M_{1},\ldots ,
UB_{r}M_{r}\} \, ;
\end{equation}
with a unitary $U$ and $(r+1)$ monomial matrices $M_{i}$ being a product of
diagonal unitaries with permutation matrices; to keep the notation simple we
do not reorder the ($r+1$) bases within $\mathcal{M}^\prime$. For the set $%
\mathcal{M}^\prime$ to be in standard form, one of the bases in $\mathcal{M}$%
, say $B_{\rho}$, must be mapped to the identity. As a consequence, the
overall unitary transformation $U$ must have a particular form, namely 
\begin{equation}  \label{U is special}
U=NB_{\rho}^{\dagger } \, ,
\end{equation}
where $N$ is some monomial matrix and $B_{\rho}$ is one of the matrices
contained in the set $\mathcal{M}$. 

In view of Eq. (\ref{U is special}) we are lead to determine the action of $%
F_{5}^{\dagger }$ and $(H_{5}^{(1)})^{\dagger }$ on the triple $\mathcal{T}%
^{(1)}$ as well as the action of $F_{5}^{\dagger }$ and $(H_{5}^{(2)})^{%
\dagger }$ on the triple $\mathcal{T}^{(2)}$. It turns out that both triples
are \emph{invariant} under these global transformations as we have the
equivalences 
\begin{equation}
NF_{5}^{\dagger }\mathcal{T}^{(1)}\sim N \mathcal{T}^{(1)}\sim
N(H_{5}^{(1)})^{\dagger }\mathcal{T}^{(1)}\,,  \label{invariant triple 1}
\end{equation}%
and 
\begin{equation}
N F_{5}^{\dagger }\mathcal{T}^{(2)}\sim N \mathcal{T}^{(2)}\sim
N(H_{5}^{(2)})^{\dagger }\mathcal{T}^{(2)}\,.  \label{invariant triple 2}
\end{equation}%
The first equivalence in (\ref{invariant triple 1}) follows from using $%
F_{5}^{\dagger }=F_{5}P$ and Eq. (\ref{fdaggeraction2}) while the second one
also requires the identity 
\begin{equation}
(H_{5}^{(1)})^{\dagger }F_{5}=H_{5}^{(4)}M\,,  \label{fleftaction1}
\end{equation}%
with some monomial matrix $M$. The equivalences (\ref{invariant triple 2})
are derived in a similar way.

Consequently, we can always remove the effect of the matrices $%
B_{\rho}^{\dagger }$ in the global transformations (\ref{U is special})
which leaves us with 
\begin{equation}  \label{limited trfs}
\{I,F_{5},H_{5}^{(j)}\} \rightarrow \{N I M_{0},N F_{5} M_{1},N H_{5}^{(j)}
M_{2}\} \, , \quad j=1,2 \, ,
\end{equation}
where $N,M_{1}$ and $M_2$ are monomial matrices, and up to rearranging
terms. The non-zero entries of the monomial matrix $N$ must, in fact, be
fifth roots of unity but we will not need this fact\footnote{%
Assume that $N$ has a nonzero element different from a fifth root, say $%
e^{i\alpha}$. This makes it impossible to transform $\mathcal{T}^{(1)}$ into
standard form using right multiplication by monomial matrices unless the
other nonzero elements of $N$ also equal $e^{i\alpha }$. It follows that $N$
must be a permutation matrix $P$ apart from a phase factor, $N=e^{i\alpha }P$%
. Thus the matrices $M_{\rho}$ must have a common factor of $e^{-i\alpha }$
which, however, is irrelevant for the definition of MU bases}.

Using the restricted transformations shown in Eqs. (\ref{limited trfs}), the
triples $\{I,F_{5},H_{5}^{(1)}\}$ and $\{I,F_{5},H_{5}^{(2)}\}$ are
equivalent to each other only if either 
\begin{equation}  \label{poss 1}
NF_{5} =F_{5}M_{1} \mbox{ and } H_{5}^{(2)}M_{2}=NH_{5}^{(1)} \, , \\
\end{equation}
or 
\begin{equation}  \label{poss 2}
NF_{5} =H_{5}^{(2)}M_{1} \mbox{ and } F_{5}M_{2} = NH_{5}^{(1)} \, ,
\end{equation}
hold for some monomial matrices $M_1$ and $M_2$. The choice $M_0 = N^{-1} =
N^\dagger$ in Eqs. (\ref{limited trfs}) ensures that the identity will be
mapped to the identity.

Eqs. (\ref{poss 1}) will now be shown to imply the identity 
\begin{equation}  \label{N removed}
\Delta F_5 = F_5 M
\end{equation}
for some monomial matrix $M$ while $\Delta$ is a diagonal matrix with fifth
roots of unity as nonzero entries, \emph{not} proportional to the identity, $%
\Delta \neq c I, c \in \mathbb{C}$. However, Eq. (\ref{N removed}) only
holds if $\Delta$ \emph{is} a multiple of the identity. This contradiction
implies that there are no matrices $N, M_1, M_2$ such that Eqs. (%
\ref{poss 1}) hold. Since Eqs. (\ref{poss 2}) also imply Eq. (\ref{N removed}) with a (possibly different) diagonal matrix $\Delta \neq c I, c
\in \mathbb{C}$, the triples $\mathcal{T}^{(1)}$ and $\mathcal{T}^{(2)}$
cannot be equivalent.

Use $H_5^{(j)} = D^j F_5, j=1,2$, to express the second equation in (\ref%
{poss 1}) as 
\begin{equation}  \label{poss 1 new}
D^2 F_{5} M_{2} = N D F_{5} = ND N^\dagger N F_5 \equiv {\tilde D} N F_5 \, ,
\\
\end{equation}
introducing ${\tilde D} \equiv ND N^\dagger = P D P^T$. Thus, the matrix ${%
\tilde D}$ is obtained from $D$ by reordering its diagonal elements
according to the permutation $P$ defined via $N=PE$, with some unitary
diagonal matrix $E$. Combining this equation with the first one in (\ref%
{poss 1}) leads to $D^2 F_{5} M_{2} = {\tilde D} F_{5}M_{1}$, or 
\begin{equation}  \label{poss 1 new prime}
{\tilde D}^\dagger D^2 F_{5} = F_{5}M_{1} M_{2}^\dagger \\
\end{equation}
which is identical to (\ref{N removed}) upon defining $\Delta = {\tilde D}%
^\dagger D^2$ and $M = M_{1} M_{2}^\dagger$ which, as a product of two
monomial matrices, is another monomial matrix. Since no permutation of the
elements on the diagonal of $D^\dagger=\mbox{diag}(1,\omega^4, \omega,
\omega, \omega^4)$ produces the inverse of $D^2$ or a multiple thereof, we
have $\Delta \neq c I$. Using the pair (\ref{poss 2}) instead of (\ref{poss 1}) also leads to an equation of the form (\ref{N removed}) with ${\tilde D}%
^\dagger$ replaced by ${\tilde D}$ which, however, cannot be a multiple of
the inverse of $D^2$, leading again to $\Delta \neq c I$.

We now show that Eq. (\ref{N removed}) only holds if the matrix $\Delta$ is
proportional to the identity. Write the monomial matrix $M$ in (\ref{N removed}) in the form 
\begin{equation}  \label{general M}
M = P \Delta^{\prime\prime} \, ,
\end{equation}
where $P$ is a permutation matrix and $\Delta^{\prime\prime}$ is a \emph{%
diagonal} matrix with entries having modulus one only. Denoting the inverse
of $\Delta^{\prime\prime}$ by $\Delta^{\prime}$, Eq. (\ref{N removed}) takes
the form 
\begin{equation}  \label{N removed new}
\Delta F_5 \Delta^\prime = F_5 P \, .
\end{equation}
Let us write $\Delta = \mbox{diag}(\alpha, \beta, \ldots, \epsilon)$ with
phase factors $\alpha, \beta$, etc, and similarly for $\Delta^\prime$, and
consider the simplest case $P \equiv I$. Then the matrix relation (\ref{N removed new}) reads explicitly  
\begin{equation}  \label{matrix elements P=I}
\left( 
\begin{array}{ccccc}
\alpha \alpha^\prime & \alpha \beta^\prime & \alpha \gamma^\prime & \alpha
\delta^\prime & \alpha \epsilon^\prime \\ 
\beta \alpha^\prime &  &  &  & \cdot \\ 
\gamma \alpha^\prime &  &  &  & \cdot \\ 
\delta \alpha^\prime &  &  &  & \cdot \\ 
\epsilon \alpha^\prime & \cdot & \cdot & \cdot & \epsilon \epsilon^\prime
\omega%
\end{array}
\right) = \left( 
\begin{array}{ccccc}
1 & 1 & 1 & 1 & 1 \\ 
1 & \omega & \omega ^{2} & \omega ^{3} & \omega ^{4} \\ 
1 & \omega ^{2} & \omega ^{4} & \omega & \omega ^{3} \\ 
1 & \omega ^{3} & \omega & \omega ^{4} & \omega ^{2} \\ 
1 & \omega ^{4} & \omega ^{3} & \omega ^{2} & \omega%
\end{array}
,\right)
\end{equation}
The conditions resulting from the first row immediately imply that the
elements on the diagonal of $\Delta^\prime$ are all equal to $\alpha^*$, or $%
\Delta^\prime = \alpha^*I$. The conditions of the first column imply that
the matrix $\Delta$ is also a multiple of the identity, namely $\Delta =
\alpha I$. This contradicts the fact that the matrix $\Delta$ is different
from a multiple of the identity.

Let us now drop the restriction the $P=I$. The effect of $P$ acting on $F_5$
from the right is to permute its columns. The first row of $F_5 $ will not
change under this operation. Under the action of $P$, the first column will
either stay where is is or it will be mapped to one of the four others. In
the first case, we can immediately apply the argument given above to derive
a contradiction. In the second case, it it straightforward to see that a
similar argument still applies involving the first row of the matrices and
that column which is the image of the first column. Thus, all possible
choices of the monomial matrix $M$ in (\ref{N removed}) require $\Delta$ to
be a multiple of the identity---which it is not.

Finally, we consider the action of an overall complex conjugation (\ref%
{conjugation})\ on either of the triples. We find that the set of three MU
bases, $\mathcal{T}^{(1)},$ remains invariant under complex conjugation 
\begin{equation}
\mathcal{T}^{(1)\,*} =\{I,{F_{5}}^*,{H}_{5}^{(1)\, *}\}
\sim \{I,F_{5},H_{5}^{(4)}\} \\
\sim \mathcal{T}^{(1)}.
\end{equation}%
Similarly, complex conjugation maps $\mathcal{T}^{(2)}$ to itself, $%
\mathcal{T}^{(2) \,*}\sim \mathcal{T}^{(2)}$. In summary, we
have shown that the equivalence relations (\ref{overallunitary}) to (\ref%
{conjugation}) cannot transform the triple $\mathcal{T}^{(1)}$ into $%
\mathcal{T}^{(2)}$ or vice versa, i.e. these triples are inequivalent.

\end{document}